\documentclass[pra,twocolumn,superscriptaddress,nofootinbib]{revtex4}
\usepackage[a4paper, left=0.8in, right=0.8in, top=1.0in, bottom=1.0in]{geometry}

\usepackage[utf8]{inputenc}
\usepackage[english]{babel}
\usepackage[T1]{fontenc}
\usepackage{amsmath}
\usepackage{amsfonts}
\usepackage{hyperref}
\usepackage{comment}
\usepackage{multirow}
\usepackage{bm}
\allowdisplaybreaks

\usepackage{chemformula}
\usepackage{tikz}
\usetikzlibrary{shapes,snakes}
\usepackage{braket}
\usepackage{natbib}
\usepackage{amsthm}

\newtheorem{theorem}{Theorem}

\newtheorem{lemma}{Lemma}

\newcommand*\chem[1]{\ensuremath{\mathrm{#1}}}

\DeclareMathOperator*{\argmin}{arg\,min}

\DeclareMathOperator{\Tr}{tr}

\newcommand{\vct}[1]{\mathbf{#1}}
\newcommand{\mtx}[1]{\mathbf{#1}}
\newcommand{\ketbra}[2]{\lvert #1 \rangle \! \langle #2 \rvert}

\usepackage[noend]{algpseudocode}
\usepackage{algorithm,algorithmicx}

\algrenewcommand\alglinenumber[1]{\sf\scriptsize\color{blue}{#1}}
\algrenewcommand\algorithmicrequire{\textbf{Input:}}
\algrenewcommand\algorithmicensure{\textbf{Output:}}

\definecolor{Red}{HTML}{E53E30}  % "Pantone 179"
\definecolor{Green}{HTML}{00AD69}  % "Pantone 3405"
\definecolor{Blue}{HTML}{2171b5}
\definecolor{Purple}{HTML}{652F6C}  % "Pantone 520"

\begin{document}

%\title{Improving randomized quantum measurements by derandomization} % Tentative
\title{Efficient estimation of Pauli observables by derandomization}
\date{\today}
\author{Hsin-Yuan Huang}
\email{hsinyuan@caltech.edu}
\affiliation{Institute for Quantum Information and Matter, Caltech, Pasadena, CA, USA}
\affiliation{Department of Computing and Mathematical Sciences, Caltech, Pasadena, CA, USA}
\author{Richard Kueng}
\affiliation{Institute for Integrated Circuits, Johannes Kepler University Linz, Austria}
\author{John Preskill}
\affiliation{Institute for Quantum Information and Matter, Caltech, Pasadena, CA, USA}
\affiliation{Department of Computing and Mathematical Sciences, Caltech, Pasadena, CA, USA}
\affiliation{Walter Burke Institute for Theoretical Physics, Caltech, Pasadena, CA, USA}
\affiliation{AWS Center for Quantum Computing, Pasadena, CA, USA}

\begin{abstract}
%\textcolor{orange}{TBD}
We consider the problem of jointly estimating expectation values of many Pauli observables, a crucial subroutine in variational quantum algorithms.
Starting with randomized measurements, we propose an efficient derandomization procedure that iteratively replaces random single-qubit measurements with fixed Pauli measurements;
the resulting deterministic measurement procedure is guaranteed to perform at least as well as the randomized one.
In particular, for estimating any $L$ low-weight Pauli observables, a deterministic measurement on only of order $\log(L)$ copies of a quantum state suffices.
In some cases, for example when some of the Pauli observables have a high weight, the derandomized procedure is substantially better than the randomized one.
Specifically, numerical experiments highlight the advantages of our derandomized protocol over various previous methods for estimating the ground-state energies of small molecules.
%Empirical studies highlight that derandomization compares favorably with current measurement protocols that do not increase circuit depth.
\end{abstract}

\maketitle

%\section{Motivation}
\section{Introduction}

Noisy Intermediate-Scale Quantum (NISQ) devices are becoming available \cite{preskill2018nisq}. Though less powerful than fully error-corrected quantum computers, NISQ devices used as coprocessors might have advantages over classical computers for solving some problems of practical interest. For example, variational algorithms using NISQ hardware have potential applications to chemistry, materials science, and optimization \cite{peruzzo2014variational,malley2016molecular,kandala2017hardware,hempel2018quantumchemistry,huang2019near,cerezo2020variational,google2020hartree,huang2020power,bharti2021noisy}.

In a typical NISQ variational algorithm, we need to estimate expectation values for a specified set of operators $\{O_1,O_2, \dots, O_L\}$ in a quantum state $\rho$ that can be prepared repeatedly using a programmable quantum system. To obtain accurate estimates, each operator must be measured many times, and finding a reasonably efficient procedure for extracting the desired information is not easy in general.
In this paper, we consider the special case where each $O_j$ is a Pauli operator; this case is of particular interest for near-term applications.

Suppose we have quantum hardware that produces multiple copies of the $n$-qubit state $\rho$. Furthermore, for every copy, we can measure all the qubits independently, choosing at our discretion to measure each qubit in the $X$, $Y$, or $Z$ basis. We are given a list of $L$ $n$-qubit Pauli operators (each one a tensor product of $n$ Pauli matrices), and our task is to estimate the expectation values of all $L$ operators in the state $\rho$, with an error no larger than $\varepsilon$ for each operator. We would like to perform this task using as few copies of $\rho$ as possible.

If all $L$ Pauli operators have relatively low weight (act nontrivially on only a few qubits), there is a simple randomized protocol that achieves our goal quite efficiently: For each of $M$ copies of $\rho$, and for each of the $n$ qubits, we chose uniformly at random to measure $X$, $Y$, or $Z$. Then we can achieve the desired prediction accuracy with high success probability if $M =O(3^{\mathrm{w}} \log L / \epsilon^2)$, assuming that all $L$ operators on our list have weight no larger than $\mathrm{w}$ \cite{evans2019scalable,huang2020predicting}.
If the list contains high-weight operators, however, this randomized method is not likely to succeed unless $M$ is very large.

In this paper, we describe a deterministic protocol for estimating Pauli-operator expectation values that always performs at least as well as the randomized protocol, and performs much better in some cases. This deterministic protocol is constructed by \emph{derandomizing} the randomized protocol. The key observation is that we can compute a lower bound on the probability that randomized measurements on $M$ copies successfully achieve the desired error $\varepsilon$ for every one of our $L$ target Pauli operators. Furthermore, we can compute this lower bound even when the measurement protocol is partially deterministic and partially randomized; that is, when some of the measured single-qubit Pauli operators are fixed, and others are still sampled uniformly from $\{X,Y,Z\}$.

Hence, starting with the fully randomized protocol, we can proceed step-by-step to replace each randomized single-qubit measurement by a deterministic one, taking care in each step to ensure that the new partially randomized protocol, with one additional fixed measurement, has success probability at least as high as the preceding protocol. When all measurements have been fixed, we have a fully deterministic protocol. In numerical experiments, we find that this deterministic protocol substantially outperforms randomized protocols
\cite{ohliger2013efficient,elben2018randomized,huang2020predicting,paini2019approximate,hadfield2020biased-shadows}. The improvement is especially significant when the list of target observables includes operators with relatively high weight.
Further performance gains are possible by executing (at least) linear-depth circuits before measurements~\cite{izmaylov2019unitary,huggins2021efficient,crawford2021efficient,yen2020cartan}.
Such procedures do, however, require deep quantum circuits. In contrast, our protocol only requires single-qubit Pauli measurements which are more amenable to execution on near-term devices.

We provide some statistical background in Sec.~\ref{sec:stat}, explain the randomized measurement protocol in Sec.~\ref{sec:random}, and analyze the derandomization procedure in Sec.~\ref{sec:derandom}. Numerical results in Sec.~\ref{sec:numerical} show that our derandomized protocol improves on previous methods. Sec.~\ref{sec:conclusion} contains concluding remarks. Further examples and details of proofs are in the appendices.

\section{Statistical background}
\label{sec:stat}
%{\it Statistical background--}
Let $\rho$ be a fixed, but unknown, quantum state on $n$ qubits. We want to accurately predict $L$ expectation values
\begin{align}
\omega_{\ell} (\rho) = \mathrm{tr}(O_{\vct{o}_{\ell}} \rho) %= \mathrm{tr} \left(\sigma_{o_{\ell,1}} \otimes \cdots \otimes \sigma_{o_{\ell,n}} \rho  \right)
\quad
\text{for $1 \leq \ell \leq L$}, \label{eq:targets}
\end{align}
where each $O_{\vct{o}_{\ell}}=\sigma_{\vct{o}_{\ell}[1]} \otimes \cdots \otimes \sigma_{\vct{o}_{\ell}[n]}$ is a tensor product of single-qubit Pauli matrices, i.e.\ $\vct{o}_{\ell}=[\vct{o}_{\ell}[1],\ldots,\vct{o}_{\ell}[n]]$ with $\vct{o}_{\ell}[k] \in \left\{I,X,Y,Z\right\}$.
To extract meaningful information, we perform $M$ (single-shot) Pauli measurements on independent copies of $\rho$. There are $3^n$ possible measurement choices. Each of them is characterized by a full-weight Pauli string $\vct{p}_m \in \left\{X,Y,Z\right\}^n$ and produces a random string of $n$ outcome signs $\vct{q}_m \in \left\{ \pm 1 \right\}^n$.

Not every Pauli measurement $\vct{p}_m$ ($1 \leq m \leq M$) provides actionable advice about every target observable $\vct{o}_{\ell}$ ($1 \leq \ell \leq L$). The two must be compatible in the sense that the latter corresponds to a marginal of the former, i.e.\ it is possible to obtain $\vct{o}_{\ell}$ from $\vct{p}_m$ by replacing some local non-identity Paulis with $I$. If this is the case, we write $\vct{o}_{\ell} \vartriangleright \vct{p}_m$ and say that measurement $\vct{p}_m$ ``hits'' target observable $\vct{o}_{\ell}$.
For instance, $[X,I],[I,X],[X,X] \vartriangleright [X,X]$, but $[Z,I],[I,Z],[Z,Z]\not \vartriangleright [X,X]$.
We can approximate each $\omega_{\ell} (\rho)$ by
empirically averaging (appropriately marginalized) measurement outcomes that belong to Pauli measurements that hit $\vct{o}_{\ell}$:
%Outcome strings $\vct{q} \in \left\{\pm 1 \right\}^n$
\begin{equation}
\hat{\omega}_{\ell} = \frac{1}{h(\vct{o}_{\ell};[\vct{p}_1,\ldots,\vct{p}_M])} \sum_{m:\; \vct{o}_{\ell} \vartriangleright \vct{p}_m} \prod_{j:\vct{o}_{\ell}[j] \neq I} \vct{q}_m[j],\label{eq:sample-mean}
%h(\vct{o}_{\ell};\mtx{P})=& \sum_{m=1}^M \mathbf{1} \left\{ \vct{o}_{\ell} \vartriangleright \vct{p}_m \right\},
\end{equation}
where
%\emph{hitting count}
$h(\vct{o}_{\ell};\left[\vct{p}_1,\ldots,\vct{p}_M \right])=\sum_{m=1}^M \mathbf{1} \left\{\vct{o}_{\ell} \vartriangleright \vct{p}_m \right\} \in \left\{0,1,\ldots,M\right\}$
counts how many Pauli measurements hit target observable $\vct{o}_{\ell}$.

It is easy to check that each $\hat{\omega}_{\ell}$ exactly reproduces $\omega_{\ell} (\rho)$ in expectation (provided that $h(\vct{o}_{\ell};\mtx{P}) \geq 1$).
Moreover, the probability of a large deviation improves exponentially with the number of hits.
%A union bound over all deviations then implies the following statement.

\begin{lemma}
[Confidence bound]
\label{lem:statistics}
Fix $\varepsilon \in (0,1)$ (accuracy) and $1-\delta \in (0,1)$ (confidence). Suppose that Pauli observables $\mtx{O}=[\vct{o}_1,\ldots,\vct{o}_L]$ and Pauli measurements $\mtx{P}=\left[\vct{p}_1,\ldots,\vct{p}_M\right]$ are such that
\begin{equation}
\textsc{Conf}_{\varepsilon}(\mtx{O};\mtx{P}):=
\sum_{\ell=1}^L \exp \left( - \tfrac{\varepsilon^2}{2}h(\vct{o}_{\ell} ;\mtx{P})\right) \leq \frac{\delta}{2}.
\label{eq:confidence-bound}
\end{equation}
Then, the associated empirical averages \eqref{eq:sample-mean} %are guaranteed to
obey
\begin{equation}
\left| \hat{\omega}_{\ell} - \omega_{\ell} (\rho) \right| \leq \varepsilon \quad \text{for all $1 \leq \ell \leq L$}
\label{eq:accurate-prediction}
\end{equation}
with probability (at least) $1-\delta$.
\end{lemma}

See Appendix~\ref{sub:concentration} for a detailed derivation.
We call the function defined in Eq.~\eqref{eq:confidence-bound} the \emph{confidence bound}.
It is a statistically sound summary parameter that checks whether a set of Pauli  measurements ($\mtx{P}$) allows for confidently predicting a collection of Pauli observables ($\mtx{O}$) up to accuracy $\varepsilon$ each.
%Intuitively speaking, a small confidence bound~\eqref{eq:confidence-bound} implies a good Pauli estimation protocol.

\section{Randomized Pauli measurements}
\label{sec:random}
%{\it Randomized Pauli measurements--}
Intuitively speaking, a small confidence bound~\eqref{eq:confidence-bound} implies a good Pauli estimation protocol. But how should we choose our $M$ Pauli measurements ($\mtx{P}$) in order to achieve $\textsc{Conf}_\varepsilon (\mtx{O};\mtx{P}) \leq \delta/2$?
The randomized measurement toolbox \cite{ohliger2013randomized,elben2018randomized,huang2020predicting,paini2019approximate,elben2020mixed} provides a perhaps surprising answer to this question.
%Fix a single Pauli observable $\vct{o}_{\ell}$ and
Let $\mathrm{w}(\vct{o}_{\ell})$ denote the \emph{weight} of Pauli observable $\vct{o}_{\ell}$, i.e.\ the number of qubits on which the observable acts nontrivially: $\mathrm{w} (\vct{o}_{\ell}) = \sum_{k=1}^n \mathbf{1} \left\{ \vct{o}_{\ell} [k]\neq I \right\}$.
These weights capture the probability of hitting $\vct{o}_{\ell}$ with a completely random measurement string: $\mathrm{Pr}_{\vct{p}} \left[ \vct{o}_{\ell} \vartriangleright \vct{p}\right]=1/3^{\mathrm{w}(\vct{o}_{\ell})}$.
%This hitting probability is exponentially suppressed in the weight, not actual system size $n$.
In turn, a total of $M$ randomly selected Pauli measurements will on average achieve $\mathbb{E}_{\mtx{P}} [h(\vct{o}_{\ell};\mtx{P})]=M/3^{\mathrm{w}(\vct{o}_{\ell})}$ hits, regardless of the actual Pauli observable $\vct{o}_{\ell}$ in question. This insight allows us to
compute expectation values of the confidence bound~\eqref{eq:confidence-bound}:
\begin{align}
\mathbb{E}_{\mtx{P}} \left[\textsc{Conf}_\varepsilon (\mtx{O};\mtx{P})\right] =&\sum_{\ell=1}^L \left(1- \nu/ 3^{\mathrm{w}(\vct{o}_{\ell})} \right)^M
\label{eq:expected-confidence-bound},
%\leq & \textcolor{blue}{L \exp \left( - \frac{\varepsilon^2 M}{4 \max_l 3^{\mathrm{w}(\mtx{O}[:,\ell])}} \right)}, \nonumber
\end{align}
where $\nu = 1-\exp (-\varepsilon^2/2) \in (0,1)$.
Each of the $L$ terms is exponentially suppressed in $\varepsilon^2M/3^{\mathrm{w}(\vct{o}_{\ell})}$.
Concrete realizations of a randomized measurement protocol are extremely unlikely to deviate substantially from this expected behavior, see e.g.\ \cite{evans2019scalable}.
Combined with Lemma~\ref{lem:statistics}, this observation implies a powerful error bound.

\begin{theorem}[Theorem~3 in Ref.~\cite{evans2019scalable}] \label{thm:randomized}
Empirical averages~\eqref{eq:sample-mean} obtained from $M$ randomized Pauli measurements allow for $\varepsilon$-accurately predicting $L$ Pauli expectation values $\mathrm{tr} (O_{\vct{o}_1} \rho),\ldots,\mathrm{tr}(O_{\vct{o}_L} \rho)$ up to additive error $\varepsilon$ given that $M \propto \log (L) \max_{\ell} 3^{\mathrm{w}(\vct{o}_{\ell})}/\varepsilon^2$.
\end{theorem}

\begin{figure*}
\begin{tikzpicture}[baseline,scale=0.7]
\draw[->,dotted,line width = 0.5mm] (-4.5,0) -- (-3.5,0);
\draw[->,dotted,line width = 0.5mm] (3.5,0) -- (4.5,0);
\begin{scope}[xshift=-8cm,yshift=0]
%\node at (0,3.5) {\textbf{first step} ($m=1$, $k=1$)};
\draw (-3,-2) rectangle (3,2);
\foreach \x in {-3,-2.75,...,3}
{\draw[gray] (\x,-2) -- (\x,2);
}
\foreach \y in {-2,-1.75,...,2}
{\draw[gray] (-3,\y) -- (3,\y);
};
%\fill[green,opacity=0.2] (-3,-2) rectangle (-1,2);
%\fill[green,opacity=0.2] (-1,0) rectangle (-0.75,2);
\fill[Red,opacity=0.8] (-3,1.75) rectangle (-2.75,2);
\fill[blue,opacity=0.2] (-3,-2) rectangle (-2.75,1.75);
\fill[blue,opacity=0.2] (-2.75,-2) rectangle (3,2);
\node at (0,2.5) {$M$ Pauli measurements};
\node[rotate=90] at (-3.5,0) {$n$ qubits};
\node at (-2.875,-2.5) {\textcolor{Red}{$m$}};
\draw[Red,->] (-2.875,-2.35) -- (-2.875,-2.05);
\node at (3.5,2-0.125) {\textcolor{Red}{$k$}};
\draw[Red,->] (3.35,2-0.125) -- (3.05,2-0.125);
\node at (0,0) {\Large \textcolor{Blue}{$\mtx{P}$}};
\end{scope}
%%%%%
\begin{scope}[xshift=0,yshift=0]
%\node at (0,3.5) {\textbf{intermediate step}};

\draw (-3,-2) rectangle (3,2);
\foreach \x in {-3,-2.75,...,3}
{\draw[gray] (\x,-2) -- (\x,2);
}
\foreach \y in {-2,-1.75,...,2}
{\draw[gray] (-3,\y) -- (3,\y);
};
\fill[teal,opacity=0.2] (-3,-2) rectangle (-1,2);
\fill[teal,opacity=0.2] (-1,0.5) rectangle (-0.75,2);
\fill[Red,opacity=0.8] (-1,0.25) rectangle (-0.75,0.5);
\fill[blue,opacity=0.2] (-1,-2) rectangle (-0.75,0.25);
\fill[blue,opacity=0.2] (-0.75,-2) rectangle (3,2);
%\node at (0,2.5) {$M$ Pauli measurements};
%\node[rotate=90] at (-3.5,0) {$n$ qubits};
\node at (-0.875,-2.5) {\textcolor{Red}{$m$}};
\draw[Red,->] (-0.875,-2.35) -- (-0.875,-2.05);
\node at (3.5,0.375) {\textcolor{Red}{$k$}};
\draw[Red,->] (3.35,0.375) -- (3.05,0.375);
\node at (-2,0) {\Large \textcolor{teal}{$\mtx{P}^\sharp$}};
\node at (1,0) {\Large \textcolor{Blue}{$\mtx{P}$}};
%\node at (0,-3) {\textcolor{Green}{$\mtx{P}^\sharp = \begin{array}{c}[\vct{p}_1^\sharp,\ldots,\vct{p}_{m-1}^\sharp] \\
%\cup [\vct{p}_m^\sharp[1],\ldots,\vct{p}_m^\sharp[k-1]]
%\end{array}$}};
\end{scope}
%%%%
\begin{scope}[xshift=8cm,yshift=0]
%\node at (0,3.5) {\textbf{last step} ($m=M,k=n$)};
\draw (-3,-2) rectangle (3,2);
\foreach \x in {-3,-2.75,...,3}
{\draw[gray] (\x,-2) -- (\x,2);
}
\foreach \y in {-2,-1.75,...,2}
{\draw[gray] (-3,\y) -- (3,\y);
};
\fill[teal,opacity=0.2] (-3,-2) rectangle (2.75,2);
\fill[teal,opacity=0.2] (2.75,-1.75) rectangle (3,2);
\fill[Red,opacity=0.8] (2.75,-2) rectangle (3,-1.75);
%\node at (0,2.5) {$M$ Pauli measurements};
%\node[rotate=90] at (-3.5,0) {$n$ qubits};
\node at (3-0.125,-2.5) {\textcolor{Red}{$m$}};
\draw[Red,->] (3-0.125,-2.35) -- (3-0.125,-2.05);
\node at (3.5,-2+0.125) {\textcolor{Red}{$k$}};
\draw[Red,->] (3.35,-2+0.125) -- (3.05,-2+0.125);
\node at (0,0) {\Large \textcolor{teal}{$\mtx{P}^\sharp$}};
%\node at (1,0) {\Large \textcolor{blue}{$\mtx{Q}$}};
\end{scope}
\end{tikzpicture}
\caption{\emph{Illustration of the derandomization algorithm (Algorithm~\ref{alg:derandomization-main}):
}
We envision $M$ randomized $n$-qubit measurements  as a 2-dimensional array comprised of $n \times M$ Pauli labels. \textcolor{Blue}{Blue squares} are placeholders for random Pauli labels, while \textcolor{teal}{green squares} denote deterministic assignments (either $X,Y$ or $Z$).
Starting with a completely unspecified array (\emph{left}), the algorithm iteratively checks
how a concrete Pauli assignment (\textcolor{red}{red square}) affects the confidence bound~\eqref{eq:confidence-bound} averaged over all remaining assignments.
A simple update rule~\eqref{eq:update-rule}
%$\mtx{P}[1,1]\gets \mathrm{argmin}_{W \in \left\{X,Y,Z\right\}} \mathbb{E}_{\mtx{P}}\left[ \textsc{Conf}_\varepsilon (\mtx{O};\mtx{P})\vert| \mtx{P}[1,1]=W\right]$
replaces the initially random label with a deterministic assignment %$\mtx{P}^\sharp[1,1] \in \left\{X,Y,Z\right\}$
that keeps the remaining confidence bound expectation as small as possible (\emph{centre}).
Once the entire grid is traversed, no randomness is left (\emph{right}) and the algorithm outputs $M$ deterministic $n$-qubit Pauli measurements.
}
\label{fig:illustration-main}
\end{figure*}
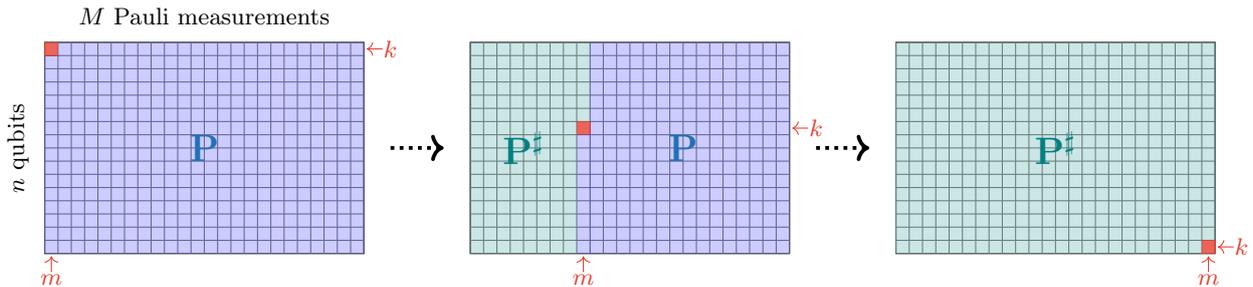

In particular, order $\log (L)$ randomized Pauli measurements suffice for estimating any collection of $L$ low-weight Pauli observables.
It is instructive to compare this result to other powerful statements about randomized measurements, most notably the ``classical shadow'' paradigm \cite{huang2020predicting,paini2019approximate}.
For Pauli observables and Pauli measurements,
the two approaches are closely related. The estimators \eqref{eq:sample-mean} are actually simplified variants of the classical shadow protocol (in particular, they don't require median of means prediction)
and the requirements on $M$ are also comparable. This is no coincidence; information-theoretic lower bounds from \cite{huang2020predicting} assert that there are scenarios where the scaling $M \propto \log (L)\max_{\ell} 3^{\mathrm{w} (\vct{o}_{\ell})} /\varepsilon^2$ is asymptotically optimal and cannot be avoided.

Nevertheless, this does not mean that randomized measurements are \emph{always} a good idea.
High-weight observables do pose an immediate challenge, because it is extremely unlikely to hit them by chance alone.

\section{Derandomized Pauli measurements}
\label{sec:derandom}
%{\it Derandomized Pauli measurements--}
The main result of this work is a procedure for identifying ``good'' Pauli measurements that allow for accurately predicting many (fixed) Pauli expectation values.
This procedure is designed to interpolate between two extremes: (i) \emph{completely randomized measurements} (good for predicting many local observables) and
(ii) \emph{completely deterministic measurements} that directly measure observables sequentially (good for predicting few global observables).

 \begin{algorithm}[t]
{\small
\begin{algorithmic}[1]
\caption{{\small \textbf{(Derandomization)} %(high-level)
}}
\label{alg:derandomization-main}

\Require
measurement budget $M$, accuracy $\varepsilon$ and $L$ Pauli observables $\mtx{O}=[\vct{o}_1,\ldots,\vct{o}_L]$

\Ensure
$M$ Pauli measurements $\mtx{P}^\sharp \in \left\{X,Y,Z\right\}^{n \times M}$
%that achieves a \textcolor{blue}{``small''} confidence bound
\vspace{0.25cm}
\Function{derandomization}{$\mtx{O},M,\varepsilon$}
\State initialize $\mtx{P}^\sharp=[]$,
%$\omega = 1 - \exp \left( -\varepsilon^2/2\right)$
\For{$m=1$ to $M$} \Comment loop of over measurements
\For{$k=1$ to $n$}\Comment loop over qubits
\For{$W=X,Y,Z$}{ \textbf{compute}} %\Comment evaluate conditional expectations
\State $f(W)~=~\mathbb{E}_{\mtx{P}}\big[ \text{Conf}_\varepsilon (\mtx{O};\mtx{P})\vert\quad\quad\quad\quad\quad\quad\quad\quad\quad\quad\quad\allowbreak \indent\indent\indent\indent\indent\indent\indent \mtx{P}^\sharp,\mtx{P}[k,m]=W\big]$
\State (see Eq.~\eqref{eq:formula} for a precise formula)
\EndFor
\State $\mtx{P}^\sharp [k,m] \gets
\text{argmin}_{W \in \left\{X,Y,Z\right\}}
f(W)$
\EndFor
\EndFor
\State output $\mtx{P}^\sharp \in \left\{X,Y,Z \right\}^{n \times M}$
\EndFunction
%%%
\end{algorithmic}
}
\end{algorithm}

Note that we can efficiently compute concrete confidence bounds~\eqref{eq:confidence-bound}, as well as expected confidence bounds averaged over all possible Pauli measurements~\eqref{eq:expected-confidence-bound}. Combined, these two formulas also allow us to efficiently compute expected confidence bounds for a list of measurements that is partially deterministic and partially randomized. Suppose that $\mtx{P}^\sharp$ subsumes deterministic assignments for the first $(m-1)$ Pauli measurements, as well as concrete choices for the first $k$ Pauli labels of the $m$-th measurement, see Fig.~\ref{fig:illustration-main} (center). Then
\begin{align}
&  \mathbb{E}_\mtx{P} \left[ \textsc{Conf}_\varepsilon (\mtx{O};\mtx{P}) \vert \mtx{P}^\sharp \right] \label{eq:formula} \\
%&=& \mathbb{E}_{\mtx{P}} \left[ \textsc{Conf}_\varepsilon (\mtx{O};\mtx{P}) \vert \mtx{P}[k',m']=\mtx{P}^\sharp [k',m'] \; \forall m' \leq m,k' \leq k, \mtx{P}[k,m]=W \right] \\
=& \sum_{\ell=1}^L
\exp \left( - \tfrac{\varepsilon^2}{2} \sum_{m'=1}^{m-1} \prod_{k'=1}^n \mathbf{1} \left\{ \vct{o}_{\ell} [k'] \vartriangleright \mtx{P}^\sharp [k',m']\right\}\right) \nonumber \\
 \times &
\left(1- \nu\prod_{k'=1}^{k}\mathbf{1} \left\{ \vct{o}_{\ell}[k'] \vartriangleright \mtx{P}^\sharp [k',m] \right\}
%3^{-\sum_{k'=k+1}^n \mathbf{1} \left\{ \vct{o}_{\ell}[k'] \neq I\right\}}
3^{-\mathrm{w}_{\neg k}(\vct{o}_{\ell})}
\right) \nonumber\\
\times & \left(1- \nu 3^{-\mathrm{w}(\vct{o}_{\ell})} \right)^{M-m},\nonumber
\end{align}
where $\nu = 1-\exp (-\varepsilon^2/2)$ and
$\mathrm{w}_{\neg k} (\vct{o}_{\ell}) = \mathrm{w} ([\vct{o}_{\ell}[k+1],\ldots,\vct{o}_{\ell}[n]])$.
This formula allows us to build deterministic measurements one Pauli-label at a time.

We start by envisioning a collection of $M$ completely random $n$-qubit Pauli measurements. That is, each Pauli label is random and Eq.~\eqref{eq:expected-confidence-bound} captures the expected confidence bound averaged over \emph{all} $3^n \times 3^M=3^{n +M}$ assignments. %Let us now focus on the first label in the first Pauli measurement for which
There are three possible choices for the first label in the first Pauli measurement: $\mtx{P}[1,1]=X$, $\mtx{P}[1,1]=Y$ and $\mtx{P}[1,1]=Z$.
At least one concrete choice does not further increase the confidence bound averaged over all remaining Pauli signs:
\begin{align}
& \min_{W \in \left\{X,Y,Z\right\}} \mathbb{E}_{\mtx{P}} \left[ \textsc{Conf}_\varepsilon (\mtx{O};\mtx{P})\vert \mtx{P}[1,1]=W \right] \label{eq:update}\\
\leq & \tfrac{1}{3}\sum_{W  \in \left\{X,Y,Z\right\}}\mathbb{E}_{\mtx{P}} \left[ \textsc{Conf}_\varepsilon (\mtx{O};\mtx{P})\vert \mtx{P}[1,1]=W \right] \nonumber\\
=& \mathbb{E}_{\mtx{P}} \left[ \textsc{Conf}_\varepsilon (\mtx{O};\mtx{P}) \right]. \nonumber
\end{align}
Crucially, Eq.~\eqref{eq:formula} allows us to efficiently identify a minimizing assignment:
\begin{equation}
\mtx{P}^\sharp[1,1]=\underset{W \in \left\{X,Y,Z\right\}}{\mathrm{argmin}}\mathbb{E}_{\mtx{P}} \left[ \textsc{Conf}_\varepsilon (\mtx{O};\mtx{P})\vert \mtx{P}[1,1]=W \right]
\label{eq:update-rule}
\end{equation}
Doing so,
replaces an initially random single-qubit measurement setting by a concrete Pauli label that minimizes the conditional expectation value over all remaining (random) assignments.
This procedure is known as %the method of conditional expectations
derandomization \cite{motwani1995randomized,spencer2008probabilistic,vazirani2001approximation} and can be iterated.
Fig.~\ref{fig:illustration-main} provides visual guidance, while pseudo-code can be found in Algorithm~\ref{alg:derandomization-main}.
There are a total of $n \times M$ iterations.
Step $(k,m)$ is contingent on comparing three conditional expectation values $\mathbb{E}_{\mtx{P}} \left[ \textsc{Conf}_\varepsilon (\mtx{O};\mtx{P})\vert \mtx{P}^\sharp, \mtx{P}[k,m]=W \right]$
and assigning the Pauli label that achieves the smallest score.
%(With a slight abuse of notation, $\mtx{P}^\sharp$ encompasses all $n(m-1) + (k-1)$ concrete assignments that have already happened up to this point.)
These update rules are constructed to ensure that
(appropriate modifications of) Eq.~\eqref{eq:update} remain valid throughout the procedure. Combining all of them implies the following rigorous statement about the resulting Pauli measurements $\mtx{P}^\sharp$.

\begin{theorem}[Derandomization promise] \label{thm:performance}
Algorithm~\ref{alg:derandomization-main}
is guaranteed to output Pauli measurements $\mtx{P}^\sharp$ with below average confidence bound:
$
\textsc{Conf}_\varepsilon (\mtx{O};\mtx{P}^\sharp) \leq \mathbb{E}_{\mtx{P}} \left[ \textsc{Conf}_\varepsilon (\mtx{O};\mtx{P}) \right]
$.
\end{theorem}

We see that derandomization produces deterministic Pauli measurements that perform at least as favorably as (averages of) randomized measurement protocols.
But the actual difference between randomized and derandomized Pauli measurements can be much more pronounced.
In the examples we considered, %a derandomized measurement protocol typically
derandomization reduces the measurement budget $M$ by at least an order of magnitude, compared to randomized measurements.
Furthermore, because Algorithm~\ref{alg:derandomization-main} implements a greedy update procedure, we have no assurance that our derandomized measurement procedure is globally optimal, or even close to optimal.

\begin{figure*}
\centering
\begin{minipage}{.48\textwidth}
    \renewcommand{\figurename}{Figure}
    \includegraphics[width=0.9\textwidth]{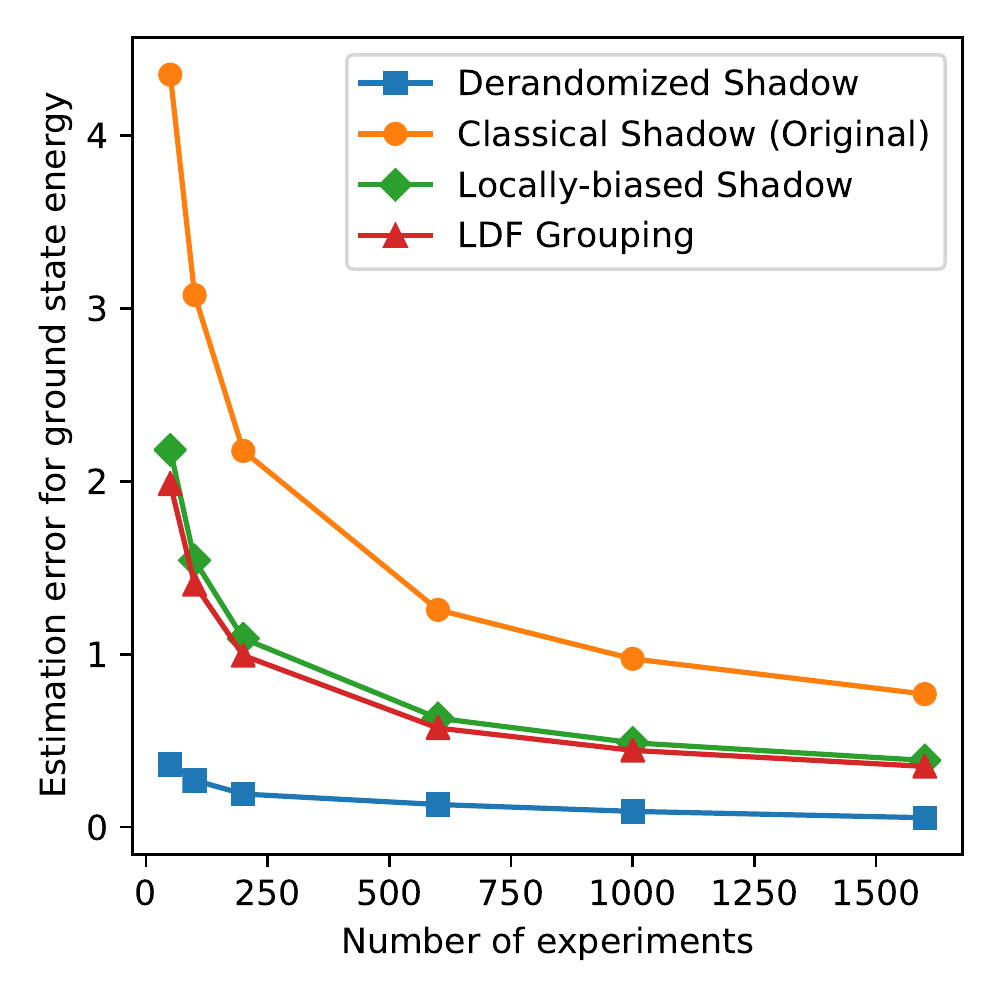}
    \caption{\emph{$\chem{Be H_2}$ ground state energy estimation error (in Hartree) under Bravyi-Kitaev encoding \cite{bravyi2002fermionic} for different measurement schemes:}
   %Estimation error of the ground state energy of $\chem{Be H_2}$ under different measurement schemes.
   The error for derandomized shadow is the root-mean-squared error (RMSE) over ten independent runs. The error for the other methods shows the RMSE over infinitely many runs and can be evaluated efficiently using the variance of one experiment \cite{hadfield2020biased-shadows}.}
    \label{fig:estimation-error}
\end{minipage}
\,\,\,\,
\begin{minipage}{.49\textwidth}
    \renewcommand{\figurename}{Table}
    \begin{tabular}{l|c|c|c|c|c}
       Molecule ($E_{\mathrm{GS}}$) & $\,$Enc.$\,$ & Derand.  & Local S. & LDF & Shadow \\
       \hline
       \multirow{3}{*}{\chem{H_2} $(-1.86)$} & JW & \textbf{0.06}  & 0.13 & 0.15 & 0.41 \\
        & P & \textbf{0.03}  & 0.14 & 0.19 & 0.48\\
       & BK & \textbf{0.06}  & 0.14 & 0.19 & 0.75 \\
       \hline
       \multirow{3}{*}{\chem{Li H} $(-8.91)$} & JW & \textbf{0.03} & 0.12 & 0.23 & 0.52 \\
       & P & \textbf{0.03}  & 0.16 & 0.29 & 0.87 \\
       & BK & \textbf{0.04} & 0.26 & 0.27 & 0.40 \\
       \hline
       \multirow{3}{*}{\chem{Be H_2} $(-19.04)$} & JW & \textbf{0.06} & 0.26 & 0.37 & 1.29\\
       & P & \textbf{0.09} & 0.36 & 0.49 & 1.77 \\
       & BK & \textbf{0.06} & 0.49 & 0.44 & 0.97 \\
       \hline
       \multirow{3}{*}{\chem{H_2 O} $(-83.60)$} & JW & \textbf{0.12} & 0.51 & 1.02 & 1.68 \\
       & P & \textbf{0.22} & 0.65 & 1.63 & 2.52 \\
       & BK & \textbf{0.20} & 1.17 & 1.45 & 3.25 \\
       \hline
       \multirow{3}{*}{\chem{N H_3} $(-66.88)$} & JW & \textbf{0.18} & 0.59 & 0.94 & 3.79 \\
       & P & \textbf{0.21} & 0.83 & 1.61 & 2.13 \\
       & BK & \textbf{0.12} & 0.73 & 1.45 & 1.89 \\
    \end{tabular}
    \caption{\emph{Average estimation error using $1000$ measurements for different molecules, encodings, and measurement schemes:} The first column shows the molecule and the corresponding ground state electronic energy (in Hartree).  We consider the following abbreviations: derandomized classical shadow (Derand.), locally-biased classical shadow (Local S.), largest degree first (LDF) heuristic and original classical shadow (Shadow) \cite{huang2020predicting}}
    \label{tab:molecules}
\end{minipage}
\end{figure*}

\section{Numerical experiments}
\label{sec:numerical}

The ability to accurately estimate many Pauli observables is an essential subroutine for variational quantum eigensolvers (VQE)~\cite{peruzzo2014variational,malley2016molecular,kandala2017ground-state,hempel2018quantumchemistry,google2020hartree}.
%%%
Randomized Pauli measurements \cite{evans2019scalable,huang2020predicting} -- also known as \emph{classical shadows} in this context -- offer a conceptually simple solution that is efficient both in terms of quantum hardware %(single-qubit Pauli measurements do not increase circuit depth)
and measurement budget.
%Further improvements are possible by adjusting the distribution of single-qubit Pauli measurements to better fit the use case at hand -- a procedure known as \emph{locally biased classical shadows} \cite{hadfield2020biased-shadows}.

\emph{Derandomization} can and should be viewed as a refinement of the original classical shadows idea. Supported by rigorous theory (Theorem~\ref{thm:performance}), this refinement is only contingent on an efficient classical pre-processing step, namely running Algorithm~\ref{alg:derandomization-main}. It does not incur any extra cost in terms of quantum hardware and classical post-processing,  but can lead to substantial performance gains.
Numerical experiments visualized in
Ref.~\cite[Figure~5]{huang2020predicting} have revealed unconditional improvements of about one order of magnitude for a particular VQE experiment \cite{kokail2019variational} (simulating quantum field theories).

In this section, we present additional numerical studies that support this favorable picture.
These address a slight variation of Algorithm~\ref{alg:derandomization-main} that does not
require fixing the total measurement budget $M$ in advance.
We focus on the \emph{electronic structure problem}: determine the ground state energy  for molecules with unknown electronic structure.
This is one of the most promising VQE applications in quantum chemistry and material science.
Different encoding shemes -- most notably \emph{Jordan-Wigner} (JW) \cite{jordan1928wigner}, \emph{Bravyi-Kitaev} (BK) \cite{bravyi2002fermionic} and \emph{Parity} (P) \cite{seeley2012parity, bravyi2002fermionic} -- allow for mapping molecular Hamiltonians to qubit Hamiltonians that correspond to sums of Pauli observables.
Several benchmark molecules have been identified whose encoded Hamiltonians are just simple enough for an explicit classical minimization, so that we can compare Pauli estimation techniques with the exact answer.

Fig.~\ref{fig:estimation-error} illustrates one such comparison. We fix a benchmark molecule \chem{Be H_2}, a Bravyi-Kitaev encoding (BK)
and plot the ground state energy approximation error against the number of Pauli measurements. The plot highlights that derandomization outperforms the original classical shadows procedure (randomized Pauli measurements) \cite{huang2020predicting}, locally-biased classical shadows \cite{hadfield2020measurements}, and another popular technique known as \emph{largest degree first (LDF) grouping} \cite{verteletskyi2020VQE-measurements,hadfield2020biased-shadows}.
The discrepancy between randomized and derandomized Pauli measurements is particularly pronounced.

This favorable picture extends to a variety of other benchmark molecules and other encoding schemes, see
Table~\ref{tab:molecules}.
For a fixed measurement budget, derandomization consistently leads to a smaller estimation error than other state-of-the-art techniques.

\section{Conclusion and outlook}
\label{sec:conclusion}

We consider the problem of predicting many Pauli expectation values from few Pauli measurements.
\emph{Derandomization} \cite{motwani1995randomized,spencer2008probabilistic,vazirani2001approximation} provides an efficient procedure that replaces originally randomized single-qubit Pauli measurements by specific Pauli assignments.
The resulting Pauli measurements are deterministic, but inherit \emph{all} advantages of a fully randomized measurement protocol.
Furthermore, the derandomization procedure could accurately capture the fine-grained structure of the observables in question.
%This makes them exceptionally well-suited for simultaneously predicting many Pauli observables with (comparatively) small weight -- an important subroutine for variational quantum eigensolvers (VQE).
%Furthermore, derandomized procedures capture the fine-grained structure of the observables in question better the original randomized measurements.
Predicting molecular ground state energies based on derandomized Pauli measurements scales favorably and improves upon many existing techniques \cite{hadfield2020biased-shadows,paini2019approximate,verteletskyi2020VQE-measurements, evans2019scalable}. Source code for an implementation of the proposed procedure is available at \cite{huangcode}.

Randomized measurements have also been used to estimate entanglement entropy \cite{brydges2019probing,huang2020predicting,vitale2021symmetry,rath2021importance}, topological invariants \cite{elben2020many,cian2021many}, benchmark physical devices \cite{knill2008randomized, elben2020mixed,huang2020predicting,choi2021emergent}, and predict outcomes of physical experiments \cite{huang2021information}.
Derandomization provides a principled approach for adapting randomized measurement procedures to fine-grained structure and is closely related to an algorithmic technique -- multiplicative weight update  \cite{arora2012multiplicative} -- commonly used in machine learning and game theory.
So far, we have only considered estimations of Pauli observables, but measurement design via derandomization should apply more broadly.
We look forward to extension of derandomization in other tasks such as estimating non-Pauli observables and entanglement entropies, as well as improvements to the cost function $f(W)$ in Algorithm~\ref{alg:derandomization-main}.

\vspace{1em}
\begin{acknowledgments}
\vspace{-1em}
The authors thank Andreas Elben, Stefan Hillmich, Steven T.\ Flammia, Jarrod McClean and Lorenzo Pastori for valuable input and inspiring discussions. HH is supported by the J. Yang \& Family Foundation.
JP acknowledges funding from  the U.S. Department of Energy Office of Science, Office of Advanced Scientific Computing Research, (DE-NA0003525, DE-SC0020290), and the National Science Foundation (PHY-1733907). The Institute for Quantum Information and Matter is an NSF Physics Frontiers Center.
\end{acknowledgments}

\bibliography{ref}
\bibliographystyle{abbrv}

\onecolumngrid
%\newpage
\appendix

%\section*{\Large Appendix}

%\section*{Appendix}

\section{Illustrative derandomization examples}

The exact workings of Algorithm~\ref{alg:derandomization-main} depend on the structure of the set of Pauli observables. In this appendix section, we provide several examples to illustrate the mechanism of the derandomization procedure.

\subsection{Many local Pauli observables.}
%{\it Example: many local Pauli observables--}
Many near-term applications of quantum devices rely on
repeatedly estimating a large number of low-weight Pauli observables. For example, low-energy eigenstates of a many-body Hamiltonian may be prepared and studied using a variational method, in which the Hamiltonian, a sum of
local terms, is measured many times.
%It is well-known that
Using randomized measurements, we can predict many low-weight observables simultaneously at comparatively little cost.
It is known that a logarithmic number of randomized Pauli measurements allows for accurately predicting a polynomial number of low-weight observables \cite{huang2020predicting}.

This desirable feature provably extends to derandomized measurements.
From Theorem~\ref{thm:performance} and Eq.~\eqref{eq:expected-confidence-bound}, we infer that the measurement budget $M=4 \log (2L/\delta) \max_{\ell} 3^{\mathrm{w}(\vct{o}_{\ell})}/\varepsilon^2$ suffices to ensure that Algorithm~\ref{alg:derandomization-main} outputs Pauli measurements $\mtx{P}^\sharp$ that obey $\textsc{Conf}_\varepsilon (\mtx{O};\mtx{P}) \leq \delta/2$. With Lemma~\ref{lem:statistics}, we may convert this into an error bound: empirical averages~\eqref{eq:sample-mean} formed from appropriate measurement outcomes are guaranteed to obey $\left| \hat{\omega}_{\ell} - \mathrm{tr}(O_{\vct{o}_{\ell}} \rho ) \right| \leq \varepsilon$ for all $1 \leq \ell \leq L$ with high probability (at least $1-\delta$).
This error bound is roughly on par with the best rigorous result about predicting local Pauli observables from randomized Pauli measurements \cite{evans2019scalable}.
But this argument implicitly assumes that $\textsc{Conf}_\varepsilon (\mtx{O};\mtx{P}^\sharp)$ (which we can compute) is comparable to $\mathbb{E}_{\mtx{P}} \left[ \textsc{Conf}_\varepsilon (\mtx{O};\mtx{P)}\right]$ (which is characterized by Eq.~\eqref{eq:expected-confidence-bound}). This assumption is extremely pessimistic, because often $\textsc{Conf}_\varepsilon (\mtx{O};\mtx{P}^\sharp) \ll \mathbb{E}_\mtx{P} \left[ \textsc{Conf}_\varepsilon (\mtx{O};\mtx{P}) \right]$.
If this is the case, derandomized Pauli measurements perform substantially better.

\subsection{Few global Pauli observables.}
%{\it Example: few global Pauli observables--}
We have seen that derandomized measurements never perform worse than randomized measurements. But they can perform much better. This discrepancy is best illustrated with a simple example: design Pauli measurements to predict both a complete $Y$-string ($\vct{o}_1=[Y,\ldots,Y])$ and a complete $Z$-string ($\vct{o}_2=[Z,\ldots,Z]$).
Here, randomized measurements are a terrible idea, because it is exponentially unlikely to hit either string by chance alone.

Contrast this with derandomization. For the very first assignment ($k=1$,$m=1$), Algorithm~\ref{alg:derandomization-main} starts by computing three conditional expectations. %$f(W)=\mathbb{E}_{\mtx{P}} \left[ \textsc{Conf}_{\varepsilon}(\mtx{O};\mtx{P}) | \mtx{P}[1,1]=W\right]$ with $W=X,Y,Z$. %$f(X)=(1-\nu/3^n)^{M-1}$, as well as $f(Y)=f(Z)=(1-\nu/3^{-(n-1)})(1-\nu/3^n)^{M-1}<f(Y)$.
Comparing them reveals $f(Y)=f(Z)<f(X)$ and the algorithm determines that assigning $X$ is likely a bad idea. The two remaining choices should be equivalent and the algorithm assigns, say, $\mtx{P}^\sharp [1,1]=Y$.
This initial choice does affect the expected confidence bound associated with the second Pauli label ($k=2$,$m=1$): %$f(Y)=(1-\nu/3^{-(n-2)})(1-\nu/3^n)^{M-1} < (1-\nu)^{-(M-1)}=f(Z)=f(X)$.
$f(Y) < f(X)=f(Z)$.
Taking into account the already assigned first Pauli label, both $X$ and $Z$ become equally unfavorable and the algorithm sticks to assigning $\mtx{P}^\sharp [2,1]=Y$. This situation now repeats itself until the first Pauli measurement is completely assigned: $\vct{p}_1^\sharp = [Y,\ldots,Y]=\vct{o}_1$. The algorithm has successfully kept track of an entire global Pauli string.

It is now time to assign the first Pauli label of the second Pauli measurement ($k=1$, $m=2$).
While $X$ is still a bad idea, taking into account that we have already measured $\vct{o}_1$ once also breaks the symmetry between $Y$ and $Z$ assignments: $f(Z) < f(Y) < f(X)$. So the algorithm chooses $\mtx{P}^\sharp [1,2]=Z$ and subsequently sticks to assigning $Z$ for all qubits: $\vct{p}^\sharp_2 = [Z,\ldots,Z]=\vct{o}_2$. Having measured both $\vct{o}_1$ and $\vct{o}_2$ an equal number of times restores the initial symmetry and the algorithm basically resets. This process resets until all $M$ Pauli measurements are assigned and Algorithm~\ref{alg:derandomization-main} outputs $\mtx{P}^{\sharp} = \left[ \vct{o}_1,\vct{o}_2,\ldots,\vct{o}_1,\vct{o}_2 \right]$.
In words: measure both global observables equally often. Although statistically optimal, this measurement protocol is neither surprising nor particularly interesting.
What is encouraging, though, is that Algorithm~\ref{alg:derandomization-main} has (re-)discovered it all by itself.

\subsection{Very many global Pauli observables (non-example):}

The derandomization algorithm is not without flaws. The greedy update rule in line 8 of Algorithm~\ref{alg:derandomization-main} can be misguided to produce non-optimal results. This happens, for instance, for a very large collection of global Pauli observables that appears to have favorable structure but actually doesn't.
For instance, set $\vct{o}_1 = [X,\ldots,X]$ and $\vct{o}_{\ell} = [Z;\tilde{\vct{o}}_{\ell}]$, where $\tilde{\vct{o}}_{\ell} \in \left\{X,Y,Z\right\}^{n-1}$ ranges through all $3^{n-1}$ possible Pauli strings of size $(n-1)$. There are $L=3^{n-1}+1$ target observables, all of which are global and therefore incompatible. However, $3^{n-1}$ of them start with a Pauli-$Z$ label.
This imbalance leads the algorithm to believe that assigning $\mtx{P}^\sharp [1,m]=Z$ for all $1 \leq m \leq M$ is always a good idea (provided that $M$ is not much larger than $3^{n-1}$). By doing so, it completely ignores the first target observable which starts with an $X$-label. But at the same time, it cannot capitalize on this particular decision, because observables $\vct{o}_2$ to $\vct{o}_{L}$ are actually incompatible.
This results in an imbalanced output $\mtx{P}^\sharp$ that treats observables $\vct{o}_2$ to $\vct{o}_L$ roughly equally, but completely forgets about $\vct{o}_1$.
Needless to say, the resulting confidence bound will not be minimal either.
We emphasize that
this highly stylized non-example is not motivated by actual applications. Instead it is intended to illustrate how greedy update procedures can get stuck in local minima.

\section{Additional details and proofs}

\subsection{Proof of Lemma~\ref{lem:statistics}} \label{sub:concentration}

Let us briefly recapitulate the general setting.
A $n$-qubit Pauli measurement $\vct{p} \in \left\{X,Y,Z\right\}^n$ produces a random string of $n$ signs $\hat{\vct{q}} \in \left\{\pm 1 \right\}^n$. Information about the underlying $n$-qubit state $\rho$ is encoded in the distribution of outcome strings
\begin{equation}
\mathrm{Pr} \left[ \hat{\vct{q}}=\vct{q} | \vct{p},\rho \right] = \mathrm{tr} \left( \bigotimes_{j=1}^m \tfrac{1}{2} \left(\sigma_I + \vct{q}[j]\sigma_{\vct{p}[j]} \right) \rho\right) \quad \text{for all $\vct{q} \in \left\{ \pm 1 \right\}^n$.}
\end{equation}
Now, suppose that $\vct{o} \in \left\{I,X,Y,Z\right\}^n$ is another Pauli string that is hit by $\vct{p}$ ($\vct{o} \vartriangleright \vct{p}$). Then, we can appropriately marginalize $n$-qubit outcome strings $\vct{q} \in \left\{ \pm 1 \right\}^n$ to reproduce $\omega(\rho) = \mathrm{tr} \left( O_{\vct{o}} \rho \right)$ in expectation:
\begin{align}
\mathbb{E} \prod_{j:\vct{o}[j]\neq I} \vct{q}[j]
=& \sum_{\vct{q} \in \left\{\pm 1\right\}^n} \mathrm{Pr} \left[ \vct{q} | \vct{p},\rho \right]
\prod_{j:\; \vct{o}_j \neq I} \vct{q}[j] \\
=& \sum_{\vct{q} \in \left\{\pm 1 \right\}^n} \mathrm{tr} \left( \bigotimes_{j: \vct{o}[j] \neq I}\tfrac{1}{2} \left( \vct{q}[j]+ \sigma_{\vct{p}[j]} \right) \bigotimes_{j:\vct{o}[j]=I} \tfrac{1}{2} \left( \sigma_I + \vct{q}[j]\sigma_{\vct{p}[j]}\right) \rho \right) \nonumber\\
=& \tfrac{1}{2^n}\sum_{\vct{q} \in \left\{ \pm 1 \right\}^n} \mathrm{tr} \left( \bigotimes_{j:\; \vct{o}[j]\neq I} \sigma_{\vct{o}[j]} \bigotimes_{j:\; \vct{o}[j]=I} \sigma_I \rho \right)
= \mathrm{tr} \left( \bigotimes_{j=1}^n\sigma_{\vct{o}[j]} \rho \right) = \mathrm{tr} \left( O_{\vct{o}} \rho \right),
\nonumber
\end{align}
whenever $\vct{o} \vartriangleright \vct{p}$ (which ensures $\vct{o}[j]=\vct{p}[j]$ whenever $\vct{o}[j]\neq I$).
Now, suppose that we perform a total of $M$ Pauli measurements $\vct{p}_1,\ldots,\vct{p}_M$.
The above relation suggests to approximate Pauli observables $\omega_{\ell} (\rho)=\mathrm{tr}(O_{\vct{o}_{\ell}} \rho )$ by empirical averages:
%Collect $\mtx{P}=[\vct{p}_1,\ldots,\vct{p}_M]$ and let $h(\vct{o}_{\ell};\mtx{P}) = \sum_{m=1}^M \mathbf{1} \left\{ \vct{o}_{\ell} \vartriangleright \vct{p}_m\right\}$ denote the hitting count.
\begin{align}
\hat{\omega}_{\ell}
= \begin{cases}
\tfrac{1}{h(\vct{o}_{\ell};\mtx{P})} \sum_{m:\vct{o}_{\ell} \vartriangleright \vct{p}_m} \prod_{j: \vct{o}_{\ell}[j]\neq I} \vct{q}_m [j] & \text{if $h(\vct{o}_{\ell};\mtx{P}) \geq 1$} \\
0 & \text{if } h(\vct{o}_{\ell};\mtx{P})=0.
\end{cases}
\label{eq:estimator-appendix}
\end{align}
Here, $h(\vct{o}_{\ell};\mtx{P}) = \sum_{m=1}^M \mathbf{1} \left\{ \vct{o}_{\ell} \vartriangleright \vct{p}_m\right\}$ denotes the \emph{hitting count}, i.e.\ the number of times a Pauli measurement $\vct{p}_m$ provides meaningful information about observable $\vct{o}_{\ell}$.
If $h(\vct{o}_{\ell};\mtx{P})=0$, not a single Pauli measurement is compatible with the target observable in question and we set $\hat{\omega}_{\ell} =0$, because we do not have any actionable advice.
The above procedure allows us to jointly estimate $L$ Pauli observables based on $M$ Pauli measurement outcomes.
The quality of reconstruction is exponentially suppressed in the number of times we hit each target Pauli observable.

\begin{lemma}
Fix a collection of $M$ Pauli measurements $\mtx{P}=[\vct{p}_1,\ldots,\vct{p}_M]$, a collection of $L$ Pauli observables $\omega_{\ell} (\rho) = \mathrm{tr} \left( O_{\vct{o}_{\ell}} \rho \right)$. Then, for all $\varepsilon >0$
\begin{align}
\mathrm{Pr} \left[\max_{1 \leq \ell \leq L} \left| \hat{\omega}_{\ell} - \omega_{\ell} (\rho) \right| \geq \varepsilon \right]
\leq 2 \sum_{\ell=1}^L\exp \left( -\tfrac{\varepsilon^2}{2}h(\vct{o}_{\ell};\mtx{P}) \right).
\end{align}
\end{lemma}

Lemma~\ref{lem:statistics} in the main text is an immediate consequence of this concentration inequality.

\begin{proof}
The union bound -- also known as Boole's inequality -- states that the probability associated with a union of events is upper bounded by the sum of individual event probabilities. For the task at hand, it implies
\begin{align}
\mathrm{Pr}\left[ \max_{1 \leq \ell \leq L}\left| \hat{\omega}_{\ell} - \omega_{\ell} (\rho) \right| \geq \varepsilon \right]
= \mathrm{Pr}\left[ \bigcup_{\ell=1}^L \left\{ \left| \hat{\omega}_{\ell} - \omega_{\ell} \right| \geq \varepsilon \right\} \right]
\leq \sum_{\ell=1}^L \mathrm{Pr} \left[ \left| \hat{\omega}_{\ell} - \omega_{\ell} (\rho) \right| \geq \varepsilon \right].
\label{eq:union-bound}
\end{align}
This allows us to treat individual deviation probabilities separately.
Fix $1 \leq \ell \leq L$ and note that $\hat{\omega}_{\ell}$ is an empirical average of $M_{\ell} = h(\vct{o}_{\ell};\mtx{P})$ random signs $s^{(\ell)}_i=\prod_{j:\;\vct{o}_{\ell}[j]\neq I} \vct{q}_i[j] \in \left\{ \pm 1 \right\}$ that are independent each (they arise from different measurement outcomes). Empirical averages of independent signed random variables tend to concentrate sharply around their true expectation value $\mathbb{E}s^{(\ell)}_i =\mathrm{tr}(O_{\vct{o}_{\ell}} \rho)$. Hoeffding's inequality makes this intuition precise and asserts for any $\varepsilon >0$
\begin{align}
\mathrm{Pr} \left[ \left| \hat{\omega}_{\ell} - \omega_{\ell} (\rho) \right| \geq \varepsilon \right]
%= \mathrm{Pr} \left[ \left| \tfrac{1}{h(\vct{o}_{\ell};\mtx{P})} \sum_{m:\vct{o}_{\ell} \vartriangleright \vct{p}_m} \left(\prod_{j:\vct{o}_{\ell}[j]\neq I} \vct{q}_m [j] - \mathbb{E} \prod_{j:\vct{o}_{\ell}[j]\neq I} \vct{q}_m [j] \right)\right|\right]
=& \mathrm{Pr} \left[\left| \tfrac{1}{M_\ell} \sum_{i=1}^{M_\ell} \left( s^{(\ell)}_i -\mathbb{E} s^{(\ell)}_i\right) \right| \geq \varepsilon \right] \leq 2 \exp \left( - \tfrac{\varepsilon^2}{2} M_{\ell} \right).
\end{align}
The claim follows, because such an exponential bound is valid for each term in Eq.~\eqref{eq:union-bound}.
This also includes terms with zero hits ($M_\ell=0$), because $\mathrm{Pr} \left[ \left| \hat{\omega}_{\ell} - \omega_{\ell} \right| \geq \varepsilon \right] \leq 1 = \exp \left( - 0/2 \right)$ --
and the claim follows.
\end{proof}

\subsection{Derivation of Eq.~\eqref{eq:formula}}

Note that each hitting count $h(\vct{o}_{\ell};\mtx{P}) = \sum_{m=1}^M \mathbf{1} \left\{ \vct{o}_{\ell} \vartriangleright \vct{p}_m \right\}$ is a sum of $M$ indicator functions that can take binary values each. This structure allows us to rewrite the confidence bound~\eqref{eq:confidence-bound} as
\begin{align}
\textsc{Conf}_{\varepsilon} (\mtx{O};\mtx{P})
=& \sum_{\ell=1}^L \exp \left( - \tfrac{\varepsilon^2}{2}h (\vct{o}_{\ell};\mtx{P})\right)
= \sum_{\ell=1}^L \prod_{m'=1}^M \exp \left( - \tfrac{\varepsilon^2}{2}\mathbf{1} \left\{ \vct{o}_{\ell} \vartriangleright \vct{p}_{m'} \right\} \right)  \label{eq:confidence-bound-app}\\
=& \sum_{\ell=1}^L \prod_{m'=1}^M \left( 1- \nu
%\left(1- \exp \left( - \tfrac{\varepsilon^2}{2}\right)\right)
\mathbf{1} \left\{ \vct{o}_{\ell} \vartriangleright \vct{p}_{m'} \right\} \right),
 \nonumber
\end{align}
where $\nu = 1- \exp \left( - \varepsilon^2/2\right) \in (0,1)$. Next, note that each remaining indicator function can be further decomposed into a product of more elementary indicator functions:
\begin{equation}
\mathbf{1} \left\{ \vct{o}_{\ell} \vartriangleright \vct{p}_{m'} \right\} = \prod_{k'=1}^n \mathbf{1} \left\{ \vct{o}_{\ell}[k']\vartriangleright \vct{p}_{m'} [k']\right\}
= \prod_{k'=1}^n \left( \mathbf{1} \left\{ \vct{o}_{\ell}[k']=I \right\} + \mathbf{1} \left\{ \vct{o}_{\ell}[k']=\vct{p}_{m'}[k']\right\} \right).
\end{equation}
Finally, note that a randomly assigned single-qubit label $\vct{p}_m[j] \in \left\{X,Y,Z\right\}$ hits non-identity Pauli label $\vct{o}_{\ell}[j]\neq I$ with probability $1/3$. More precisely,
\begin{align}
\mathbb{E}_{\vct{p}_m[j]}
\left[\mathbf{1} \left\{ \vct{o}_{\ell}[j] \vartriangleright \vct{p}_m [j] \right\} \right]
= \mathrm{Pr}_{\vct{p}_m[j]} \left[ \vct{o}_{\ell}[j] \vartriangleright \vct{p}_m [j] \right]
= (1/3)^{\mathbf{1} \left\{ \vct{o}_{\ell}[j]\neq I \right\}}
=
\begin{cases}
1/3 & \text{if $\vct{o}_{\ell}[j]\neq I$,} \\
1 & \text{if $\vct{o}_{\ell}[j]=I$.}
\end{cases}
\end{align}
Together with independence, this observation allows us to compute expectation values of confidence bounds that are partially assigned already.
Let $\mtx{P}^\sharp$ denote the already assigned part that  encompasses the first $m-1$ Pauli measurements, as well as the first $k$ single-qubit labels of the $m$-th Pauli measurement: $\mtx{P}^\sharp = \left[\vct{p}_1^\sharp,\ldots,\vct{p}_{m-1}^\sharp \right] \cup \left[ \vct{p}_m^\sharp [1],\ldots,\vct{p}_m[k]^\sharp \right]$.
We also assume that all remaining Pauli labels are assigned independently and uniformly at random ($\mathrm{Pr} \left[ \vct{p}_{m'}[k'] = X\right] = \mathrm{Pr} \left[ \vct{p}_{m'}[k']=Y\right]=\mathrm{Pr} \left[ \vct{p}_{m'}[k']=Z\right]=1/3$). Independence ensures that the conditional expectation factorizes nicely into individual components:
\begin{align}
\mathbb{E}_{\mtx{P}} \left[ \textsc{Conf}_{\varepsilon}(\mtx{O};\mtx{P})\rvert \mtx{P}^\sharp \right]
%=& \mathbb{E}_{\mtx{P}} \left[ \textsc{Conf}_{\varepsilon}(\mtx{O};\mtx{P})\rvert \vct{p}_1=\vct{p}_1^\sharp,\ldots,\vct{p}_{m-1}=\vct{p}_{m-1}^\sharp,\vct{p}_m[1]=\vct{p}_m[1]^\sharp,\ldots,\vct{p}_m[k]=\vct{p}_m^\sharp [k] \right] \nonumber\\
=& \sum_{\ell=1}^L \prod_{m'=1}^{m-1} \left( 1- \nu \mathbf{1} \{ \vct{o}_{\ell} \vartriangleright \mtx{p}_{m'}^\sharp \} \right) \\
\times & \left( 1- \nu \prod_{k'=1}^k \{ \vct{o}_{\ell}[k'] \vartriangleright \vct{p}_m [k']\} \prod_{k'=k+1}^n\mathbb{E}_{\vct{p}_m[k']} \{ \vct{o}_{\ell}[k'] \vartriangleright \vct{p}_m[k']\} \right) \nonumber \\
\times & \prod_{m'=m+1}^M \left(1- \nu \prod_{k'=1}^n \mathbb{E}_{\vct{p}_{m'}[k']}\mathbf{1} \{ \vct{o}_{\ell}[k'] \vartriangleright \vct{p}_{m'}[k'] \} \right) \nonumber\\
=& \sum_{\ell=1}^L \prod_{m'=1}^{m-1}\left(1- \nu \mathbf{1} \{ \vct{o}_{\ell} \vartriangleright \vct{p}_{m'}^\sharp \} \right)
\left( 1- \nu \prod_{k'=1}^k \{ \vct{o}_{\ell}[k'] \vartriangleright \vct{p}_m [k'] \} \prod_{k'=k+1}^n (1/3)^{\mathbf{1} \left\{ \vct{o}_{\ell}[k']\neq I\right\}} \right) \nonumber \\
\times & \prod_{m'=m+1}^M \left( 1 - \nu \prod_{k'=1}^n (1/3)^{\mathbf{1} \left\{ \vct{o}_{\ell}[k'] \neq I \right\}} \right). \nonumber
\end{align}
Now, note that the exponent
$\sum_{k'=k+1}^n \mathbf{1}\{\vct{o}_{\ell}[k']\neq I\}=\mathrm{w}_{\neg k}(\vct{o}_{\ell})$ captures the weight of the reduced Pauli string $[\vct{o}_{\ell}[k+1],\ldots,\vct{o}_{\ell}[n]]$ (in particular, $\mathrm{w}_{\neg 0} (\vct{o}_{\ell}) = \mathrm{w} (\vct{o}_{\ell})$)
Reading Eq.~\eqref{eq:confidence-bound-app} backwards to recognize $\prod_{m'=1}^{m-1} \left(1-\nu \mathbf{1} \{\vct{o}_{\ell} \vartriangleright \vct{p}_{m'}^\sharp \} \right)=\exp \left( - \tfrac{\varepsilon^2}{2}h(\vct{o}_{\ell};[\vct{p}_1^\sharp,\ldots,\vct{p}_{m-1}^\sharp ])\right)$
further simplifies the expression:
\begin{align}
\mathbb{E}_{\mtx{P}}\left[ \textsc{Conf}_{\varepsilon}(\mtx{O};\mtx{P}^\sharp)\vert \mtx{P}^\sharp \right]
=& \sum_{\ell=1}^L \exp \left( - \tfrac{\varepsilon^2}{2} h(\vct{o}_{\ell};[\vct{p}_1^\sharp,\ldots,\vct{p}_{m-1}^\sharp])\right) \left( 1- \nu \prod_{k'=1}^k \{ \vct{o}_{\ell}[k'] \vartriangleright \vct{p}_m[k']\} 3^{-\mathrm{w}_{\neg k}(\vct{o}_{\ell})} \right) \\
\times &\left( 1- \nu 3^{-\mathrm{w}(\vct{o}_{\ell})} \right)^{M-m}. \nonumber
\end{align}

\section{Details regarding numerical experiments} \label{app:numerics}

We consider a molecular electronic Hamiltonian that has been encoded into an $n$-qubit system.
The Hamiltonian can be written as a sum of Pauli observables.
\begin{equation} \label{eq:Hpauli}
    H = \sum_{P \in \{I, X, Y, Z\}^{n}} \alpha_P P.
\end{equation}
The number of qubits for different molecules is given by
\begin{equation}
    \chem{H_2}: n=8, \,\, \chem{Li H}: n=12, \,\, \chem{Be H_2}: n=14, \,\, \chem{H_2 O}: n=14, \,\, \chem{N H_3}: n=16.
\end{equation}
Each molecule is represented by a fermionic Hamiltonian in a minimal STO-$3$G basis, ranging from $4$ to $16$ spin orbitals. The $8$-qubit $\chem{H_2}$ example is represented using a $6$-$31$G basis.
The fermionic Hamiltonian is mapped to a qubit Hamiltonian using three different common encodings: \emph{Jordan-Wigner} (JW) \cite{jordan1928wigner}, \emph{Bravyi-Kitaev} (BK) \cite{bravyi2002fermionic} and \emph{Parity} (P) \cite{seeley2012parity, bravyi2002fermionic}.
The Pauli decomposition considered here has already been featured in many existing works; see \cite{kandala2017hardware, bravyi2017tapering, hadfield2020measurements} for more details.

%The measurement procedures require a state $\rho$ to be measured.
In our numerical experiments, the measurement procedure is applied to the exact ground state of the encoded $n$-qubit Hamiltonian $H$:
\begin{equation}
    \rho = \ketbra{g}{g},\quad \text{where} \quad \ket{g} = \argmin_{\ket{\psi}} \bra{\psi}H\ket{\psi}.
\end{equation}
The ground state $\ket{g}$ is obtained by exact diagonalization using the Lanczos method, see e.g.\ \cite{tropp2020randomized} for a recent survey.
We focus on root-mean squared error (RMSE) to quantify the measurement error.
For $M$ independent repetitions of the measurement procedure giving rise to $M$ estimates $\hat{E}_1, \ldots, \hat{E}_M$, the RMSE is given by:
\begin{equation}
    \mathrm{RMSE} = \sqrt{\frac{1}{M} \sum_{i=1}^M (\hat{E}_i - E_{\mathrm{GS}})^2},
\end{equation}
where $E_{\mathrm{GS}}$ is the exact ground state electronic energy $\Tr(H \rho) = \bra{\psi}H\ket{\psi}$.
We consider the ground state electronic energy of the molecule without the static Coulomb repulsion energy between the nuclei. Hence the total ground state energy of the molecule is the sum of the ground state electronic energy and the static Coulomb repulsion energy (Born-Oppenheimer approximation).
We do not focus on the static Coulomb repulsion energy because it is not encoded in the molecular electronic Hamiltonian $H$ and is considered to be a fixed value.

We elaborate the alternative measurement procedures with which we compared our derandomized procedure.
\begin{enumerate}
    \item \emph{LDF grouping}: The largest-degree-first (LDF) grouping strategy and other heuristics have been considered and investigated in \cite{verteletskyi2020measurement}.
    The conclusion is that the LDF grouping strategy results in good performance (differing from the best heuristics by at most $10\%$) and is generally recommended.
    The measurement error (RMSE) of LDF grouping strategy can be computed exactly given an exact representation of the ground state $\ket{g}$; see \cite{hadfield2020measurements} for details.
    \item \emph{Classical shadow}: The measurement procedure measures each qubit in a random $X, Y, Z$ Pauli basis.  This procedure is known to allow estimation of any $L$ few-body observables from only order $\log(L)$ measurements \cite{cotler2020quantum, evans2019scalable, huang2020predicting}. However, the performance would degrade significantly when we consider many-body observables. Hence, this approach will likely perform less well for molecular Hamiltonians due to the presence of many high-weight Pauli observables.
    \item \emph{Locally-biased classical shadow}: This is an improvement over classical shadows, proposed by \cite{hadfield2020measurements}, designed to overcome disadvantages in estimating the expectation of many-body observables. The idea is to bias the distribution over different Pauli bases ($X,Y$ or $Z$) for each qubit to minimize the variance when we measure the quantum Hamiltonian given in Equation~\eqref{eq:Hpauli}. Ref.~\cite{hadfield2020measurements} demonstrated that this approach would yield similar or better performance compared to LDF grouping and outperforms classical shadows.
\end{enumerate}

In what follows, we provide a detailed description of the cost function used to derandomize the single-qubit Pauli observables for our numerical experiments. In Algorithm~\ref{alg:derandomization-main}, we used the cost function
\begin{equation}
    f(W) = \mathbb{E}_{\mtx{P}}\big[ \text{Conf}_\varepsilon (\mtx{O};\mtx{P})\vert  \mtx{P}^\sharp,\mtx{P}[k,m]=W\big].
\end{equation}
The conditional expectation is given by Eq.~\eqref{eq:formula} and is restated here for convenience
\begin{align}
\mathbb{E}_\mtx{P} \left[ \textsc{Conf}_\varepsilon (\mtx{O};\mtx{P}) \vert \mtx{P}^\sharp \right]
%&=& \mathbb{E}_{\mtx{P}} \left[ \textsc{Conf}_\varepsilon (\mtx{O};\mtx{P}) \vert \mtx{P}[k',m']=\mtx{P}^\sharp [k',m'] \; \forall m' \leq m,k' \leq k, \mtx{P}[k,m]=W \right] \\
=& \sum_{\ell=1}^L
\exp \left( - \frac{\varepsilon^2}{2} \sum_{m'=1}^{m-1} \prod_{k'=1}^n \mathbf{1} \left\{ \vct{o}_{\ell} [k'] \vartriangleright \mtx{P}^\sharp [k',m']\right\}\right) \nonumber \\
 \times &
\left(1- \nu\prod_{k'=1}^{k}\mathbf{1} \left\{ \vct{o}_{\ell}[k'] \vartriangleright \mtx{P}^\sharp [k',m] \right\}
%3^{-\sum_{k'=k+1}^n \mathbf{1} \left\{ \vct{o}_{\ell}[k'] \neq I\right\}}
3^{-\mathrm{w}_{\neg k}(\vct{o}_{\ell})}
\right) \nonumber\\
\times & \left(1- \nu 3^{-\mathrm{w}(\vct{o}_{\ell})} \right)^{M-m},\nonumber
\end{align}
where $\nu = 1-\exp (-\varepsilon^2/2)$ and
$\mathrm{w}_{\neg k} (\vct{o}_{\ell}) = \mathrm{w} ([\vct{o}_{\ell}[k+1],\ldots,\vct{o}_{\ell}[n]])$.
This formula requires us to fix the total number of measurements $M$ beforehand. However, one may want to keep measuring until certain criteria are satisfied, e.g., that all of the $L$ Pauli observables has been measured sufficiently many times.
In such a scenario, it is unclear what $M$ should be.
One approach is to try out various different values of $M$ and choose the one that works best.
In the numerical experiments, we consider the following alternative strategy, where we simply remove $\left(1- \nu 3^{-\mathrm{w}(\vct{o}_{\ell})} \right)^{M-m}$ since it only depends on the weight of the Pauli observable $\vct{o}_{\ell}$.
The results are similar and one does not have to choose $M$ beforehand.
The precise formula we used in Algorithm~\ref{alg:derandomization-main} is now given by a modified cost function instead of the conditional expectation value,
\begin{equation}
    f(W) = C(\mtx{P}^\sharp,\mtx{P}[k,m]=W).
\end{equation}
The modified cost function is a sum of single-observable cost functions $\exp \left( - V(\vct{o}_\ell, \mtx{P}^\sharp) \right)$,
\begin{align}
    C(\mtx{P}^\sharp) &= \sum_{\ell=1}^L
\exp \left( - V(\vct{o}_\ell, \mtx{P}^\sharp) \right),\\
V(\vct{o}_\ell, \mtx{P}^\sharp) &=  \frac{\eta}{2} \sum_{m'=1}^{m-1} \prod_{k'=1}^n \mathbf{1} \left\{ \vct{o}_{\ell} [k'] \vartriangleright \mtx{P}^\sharp [k',m']\right\} \nonumber \\
& -
\log\left(1- \frac{\nu}{3^{\mathrm{w} ([\vct{o}_{\ell}[k+1],\ldots,\vct{o}_{\ell}[n]])}} \prod_{k'=1}^{k}\mathbf{1} \left\{ \vct{o}_{\ell}[k'] \vartriangleright \mtx{P}^\sharp [k',m] \right\}
\right), \label{def:Vcost}
\end{align}
where $\eta, \nu > 0$ are hyperparameters that need to be chosen properly. In the numerical experiments, we consider $\eta = 0.9$ and $\nu = 1-\exp (-\eta / 2)$. The larger $V(\vct{o}_\ell, \mtx{P}^\sharp)$ is, the lower the single-observable cost function $\exp \left( - V(\vct{o}_\ell, \mtx{P}^\sharp) \right)$ will be. The following discussion provides an intuitive understanding for the role of the two terms in $V(\vct{o}_\ell, \mtx{P}^\sharp)$.
\begin{enumerate}
    \item The first term in $V(\vct{o}_\ell, \mtx{P}^\sharp)$ is proportional to
    \begin{equation}
        \sum_{m'=1}^{m-1} \prod_{k'=1}^n \mathbf{1} \left\{ \vct{o}_{\ell} [k'] \vartriangleright \mtx{P}^\sharp [k',m']\right\},
    \end{equation}
    which determines how many times the Pauli observable $\vct{o}_{\ell}$ has been measured in the first $m-1$ Pauli measurements.
    If the Pauli observable $\vct{o}_{\ell}$ has been measured many times, then $V(\vct{o}_\ell, \mtx{P}^\sharp)$ is large, and therefore $\exp \left(- V(\vct{o}_\ell, \mtx{P}^\sharp) \right)$ is close to zero.
    \item The second term in $V(\vct{o}_\ell, \mtx{P}^\sharp)$ is approximately equal to the following by Taylor expansion,
    \begin{equation}
        \frac{\nu}{3^{\mathrm{w} ([\vct{o}_{\ell}[k+1],\ldots,\vct{o}_{\ell}[n]])}} \prod_{k'=1}^{k}\mathbf{1} \left\{ \vct{o}_{\ell}[k'] \vartriangleright \mtx{P}^\sharp [k',m] \right\}.
    \end{equation}
    It would be nonzero only when $\vct{o}_{\ell}[k'] \vartriangleright \mtx{P}^\sharp [k',m]$ for all $k' = 1, \ldots, k$. Furthermore if the weight of $[\vct{o}_{\ell}[k+1],\ldots,\vct{o}_{\ell}[n]]$ is smaller, then the single-observable cost function $\exp \left( - V(\vct{o}_\ell, \mtx{P}^\sharp) \right)$ incurred by $\vct{o}_\ell$ would be smaller.
\end{enumerate}
When the entire set of $M$ measurements has been decided, $V(\vct{o}_\ell, \mtx{P}^\sharp)$ will consist only of the first term and is proportional to the number of times the observable $\vct{o}_\ell$ has been measured.

For quantum chemistry applications, the coefficients of different Pauli observable are different, e.g., in Eq.~\eqref{eq:Hpauli}, the Hamiltonian $H$ consists of Pauli observable $P$ with varying coefficients $\alpha_P$.
In such a case, one would want to measure each Pauli observable $\vct{o}_\ell$ with a number of times proportional to $|\alpha_{\vct{o}_\ell}|$ \cite{mcclean2016theory}.
In order to include the proportionality to $|\alpha_{\vct{o}_\ell}|$, we consider the following modified cost function that depends on the coefficients $\alpha$,
\begin{equation}
     C_{\alpha}(\mtx{P}^\sharp) = \sum_{l=1}^L \exp\left( - V(\vct{o}_\ell, \mtx{P}^\sharp) / w_{\vct{o}_\ell} \right), \,\,\, \mbox{where} \,\,\, w_{\vct{o}_\ell} = \frac{|\alpha_{\vct{o}_\ell}|}{\max_{p} |\alpha_{\vct{o}_{p}}|}.
\end{equation}
The definition of $V(\vct{o}_\ell, \mtx{P}^\sharp)$ is given in Eq.~\eqref{def:Vcost}.
Recall that $V(\vct{o}_\ell, \mtx{P}^\sharp)$ will be proportional to the number of times the observable $\vct{o}_\ell$ has been measured, hence the weight factor $w_{\vct{o}_\ell}$ will promote the proportionality of $V(\vct{o}_\ell, \mtx{P}^\sharp)$ to $w_{\vct{o}_\ell} \propto |\alpha_{\vct{o}_\ell}|$.
While the cost function is derived from derandomizing the powerful randomized procedure \cite{huang2020predicting}, it is not clear if this is the optimal cost function. We believe other cost functions that are tailored to the particular application could yield even better performance; we leave such an exploration as goal for future work.

\end{document}